\documentclass[]{spie}  %>>> use for US letter paper
\usepackage[]{graphicx}
\usepackage{amsmath}
\usepackage{gensymb}
\usepackage{upgreek}

\def\apj{ApJ}%
\def\aap{A\&A}%
\def\pasp{PASP}%
\def\procspie{SPIE proceedings}

\title{Simultaneous Water Vapor and Dry Air Optical Path Length Measurements and Compensation with the Large Binocular Telescope Interferometer} 

\author{D.~Defr\`ere\supit{a,b}, P.~Hinz\supit{a}, E.~Downey\supit{a}, M.~B\"ohm\supit{c}, W.C.~Danchi\supit{d}, O.~Durney\supit{a}, S.~Ertel\supit{a}, J.M.~Hill\supit{e}, W.F.~Hoffmann\supit{a}, B.~Mennesson\supit{f}, R.~Millan-Gabet\supit{g}, M.~Montoya\supit{a}, J.-U.~Pott\supit{h}, A.~Skemer\supit{i}, E.~Spalding\supit{a}, J.~Stone\supit{a}, A.~Vaz\supit{a}
\skiplinehalf
\supit{a}Steward Observatory, University of Arizona, 933 N. Cherry Avenue, 85721 Tucson, USA\\
\supit{b}STAR Institute, Universit\'e de Li\`ege, 17 All\'ee du Six Ao\^ut, B-4000 Sart Tilman, Belgium\\
\supit{c}ISYS, University of Stuttgart, Pfaffenwaldring 9, 70569 Stuttgart, Germany\\
\supit{d}NASA Goddard Space Flight Center, Exoplanets \& Stellar Astrophysics Laboratory, Code 667, Greenbelt, MD 20771\\
\supit{e}Large Binocular Telescope Observatory, University of Arizona, 933 N. Cherry Avenue, 85721 Tucson, USA\\
\supit{f}Jet Propulsion Laboratory, California Institute of Technology 4800 Oak Grove Drive, Pasadena CA 91109-8099, USA\\
\supit{g}NASA Exoplanet Science Center (NExSci), California Institute of Technology, 770 South Wilson Avenue, Pasadena CA 91125, USA\\
\supit{h}Max-Planck-Institute for Astronomy, K\"onigstuhl 17, 69117 Heidelberg, Germany\\
\supit{i}Department of Astronomy and Astrophysics, University of California, Santa Cruz, 1156 High St, Santa Cruz, CA 95064, USA\\
}

\authorinfo{Send email correspondence to D.D. at denis@lbti.org.}
\begin{document} 
  \maketitle 

%%%%%%%%%%%%%%%%%%%%%%%%%%%%%%%%%%%%%%%%%%%%%%%%%%%%%%%%%%%%% 
\begin{abstract}
The Large Binocular Telescope Interferometer uses a near-infrared camera to measure the optical path length variations between the two AO-corrected apertures and provide high-angular resolution observations for all its science channels (1.5-13~$\upmu$m). There is however a wavelength dependent component to the atmospheric turbulence, which can introduce optical path length errors when observing at a wavelength different from that of the fringe sensing camera. Water vapor in particular is highly dispersive and its effect must be taken into account for high-precision infrared interferometric observations as described previously for VLTI/MIDI or the Keck Interferometer Nuller. In this paper, we describe the new sensing approach that has been developed at the LBT to measure and monitor the optical path length fluctuations due to dry air and water vapor separately. After reviewing the current performance of the system for dry air seeing compensation, we present simultaneous H-, K-, and N-band observations that illustrate the feasibility of our feedforward approach to stabilize the path length fluctuations seen by the LBTI nuller.  
\end{abstract}

%>>>> Include a list of keywords after the abstract 

%%%%%%%%%%%%%%%%%%%%%%%%%%%%%%%%%%%%%%%%%%%%%%%%%%%%%%%%%%%%%
\section{INTRODUCTION}
\label{sec:intro}  % \label{} allows reference to this section

Water vapor turbulence can limit the performance of high dynamic range interferometers that use phase-referenced modes, even though such turbulence is only a small contributor to the total seeing at visible and infrared wavelengths. Indeed, because water vapor is highly dispersive, random fluctuations in its differential column density above each aperture (or water vapor seeing) will create a chromatic component to the optical path difference (OPD) that is not properly tracked at wavelengths different from that of the fringe sensor. The impact of this effect on infrared interferometry has been addressed extensively in the literature, either in a general context\cite{Colavita:2004} or applied to specific instruments that include phase-referenced modes using K-band light such as VLTI/MIDI\cite{Meisner:2003,Matter:2010,Pott:2012,Muller:2014}, VLTI/GENIE\cite{Absil:2006}, or the Keck Interferometer Nuller\cite{Koresko:2003,Koresko:2006,Colavita:2010} (KIN). In this paper, we address the effect of precipitable water vapor (PWV) in the context of the Large Binocular Telescope Interferometer\cite{Hinz:2008} (LBTI), which uses a K-band fringe sensor to provide high-bandwidth path length compensation for all its scientific channels (1.5-13\,$\upmu$m) and in particular for the nulling interferometer that operates at 11\,$\upmu$m. 

 %We present in Section~\ref{sec:model} the model predictions for the LBTI
%\cite{Defrere:2015}
%\cite{Defrere:2015b}

\section{PREDICTIONS FOR THE LBTI}
\label{sec:model}

Because both apertures are installed on a single steerable mount, each beam travels almost exactly the same path length in both the atmosphere and the LBTI cryostat. Therefore, the deterministic dispersion due to the differential path length in humid air, which impacts most long-baseline interferometers, can be ignored in the case of the LBT and we will focus the following analysis on random atmospheric dispersion due to water vapor seeing (which is the dominant contributor). According to previous studies\cite{Meisner:2003,Colavita:2004}, the random component due to water vapor density variations are expected to be about 10$^{-2}$ of the total water vapor in the air column, which is 3 orders of magnitude larger than the equivalent value for dry air density variations (i.e., less than 10$^{-5}$ of the total air column). Dry air still dominates the refractivity and seeing because it is the dominant volume contributor but atmospheric water vapor is much more dispersive, with a ratio of water vapor to dry air dispersion of 200:1 at K band\cite{Colavita:2004}. Therefore, variable differential water-vapor column densities above each aperture will create random and wavelength-dependent OPD variations (on longer timescales than dry air, typically a few seconds to a few minutes). Measurements with the KIN have shown that this effect produces an OPD RMS error of 0.32\,$\upmu$m at N band when tracking the fringes at K band in median conditions at Mauna Kea\cite{Colavita:2010} (a factor 2.3 smaller than theoretical predictions). For the LBTI, we expect a similar value due to the combination of two opposing effects: the wetter median conditions at Mount Graham and the shorter baseline (14.4\,m vs 82\,m), which sees more correlated piston turbulence and hence less OPD variations (depending on the outer scale of the water vapor turbulence). 
%
%\begin{figure}[!t]
%	\begin{center}
%		\includegraphics[width=16.5 cm]{block_diagram.eps}
%		\caption{System-level block diagram of LBTI architecture in nulling mode showing the optical path through the telescope, beam combiner (red box), and the NIC cryostat (blue box). After being reflected on LBT primaries, secondaries, and tertiaries, the visible light is reflected on the entrance window and used for wavefront sensing while the infrared light is transmitted into LBTI, where all subsequent optics are cryogenic. The beam combiner directs the light with steerable mirrors and can adjust pathlength for interferometry. Inside the NIC cryostat, the thermal near-infrared (3-5 $\upmu$m) light is directed to LMIRCam for exoplanet imaging, the near-infrared (1.5-2.5 $\upmu$m) light is directed to the phase sensor, which measures the differential tip/tilt and phase between the two primary mirrors, and the mid-infrared (8-13 $\upmu$m) light is directed to NOMIC for nulling interferometry. Both outputs of the beam combiner are directed to the phase and tip/tilt sensor, while only the nulled output of the interferometer is reflected to the NOMIC camera with a short-pass dichroic. The various cameras are shown in dark grey and feed-back signals driving the deformable secondary mirrors and tip-tilt/OPD correctors are represented by dashed lines. Note that this diagram is schematic only and does not show several additional optics.}\label{fig:diagram}
%	\end{center}
%\end{figure}

\section{MEASUREMENT APPROACH}
\label{sec:measurement}  
\subsection{Dry air measurements}

\begin{figure}[!t]
	\begin{center}
	        \includegraphics[height=5.0 cm]{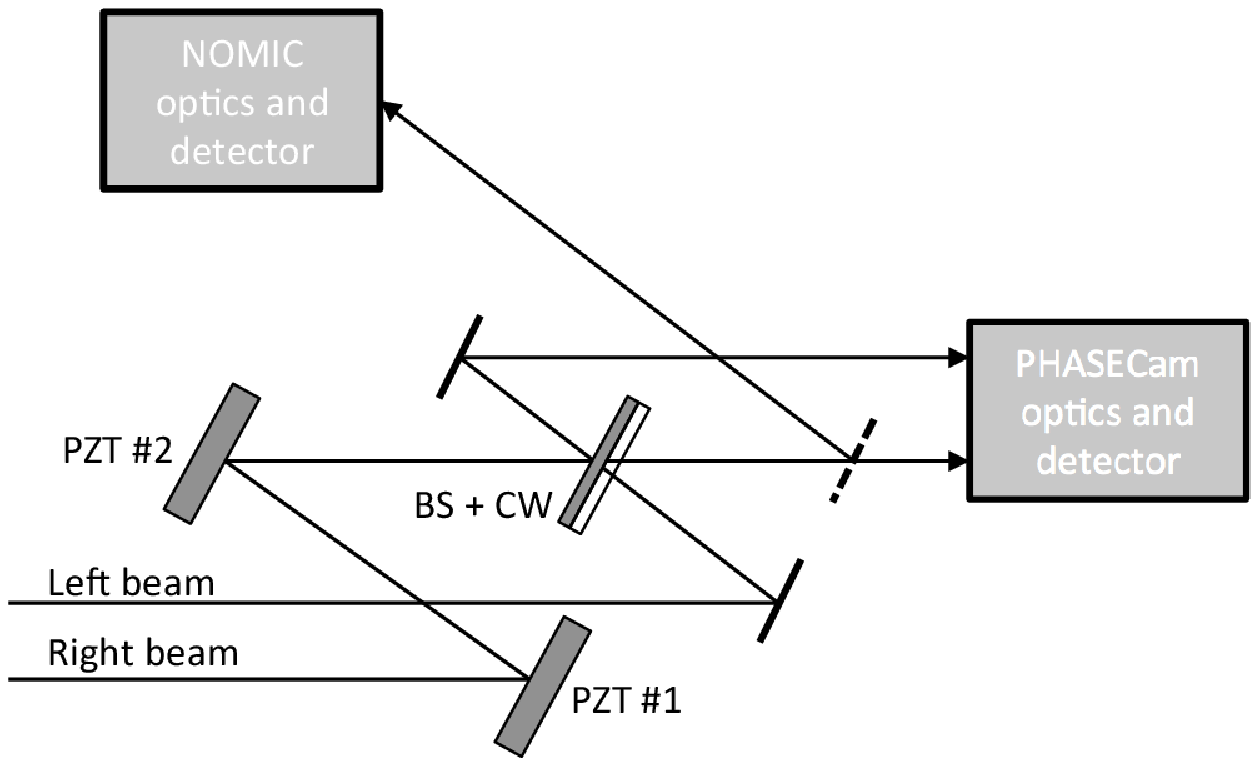}
		\includegraphics[height=5.0 cm]{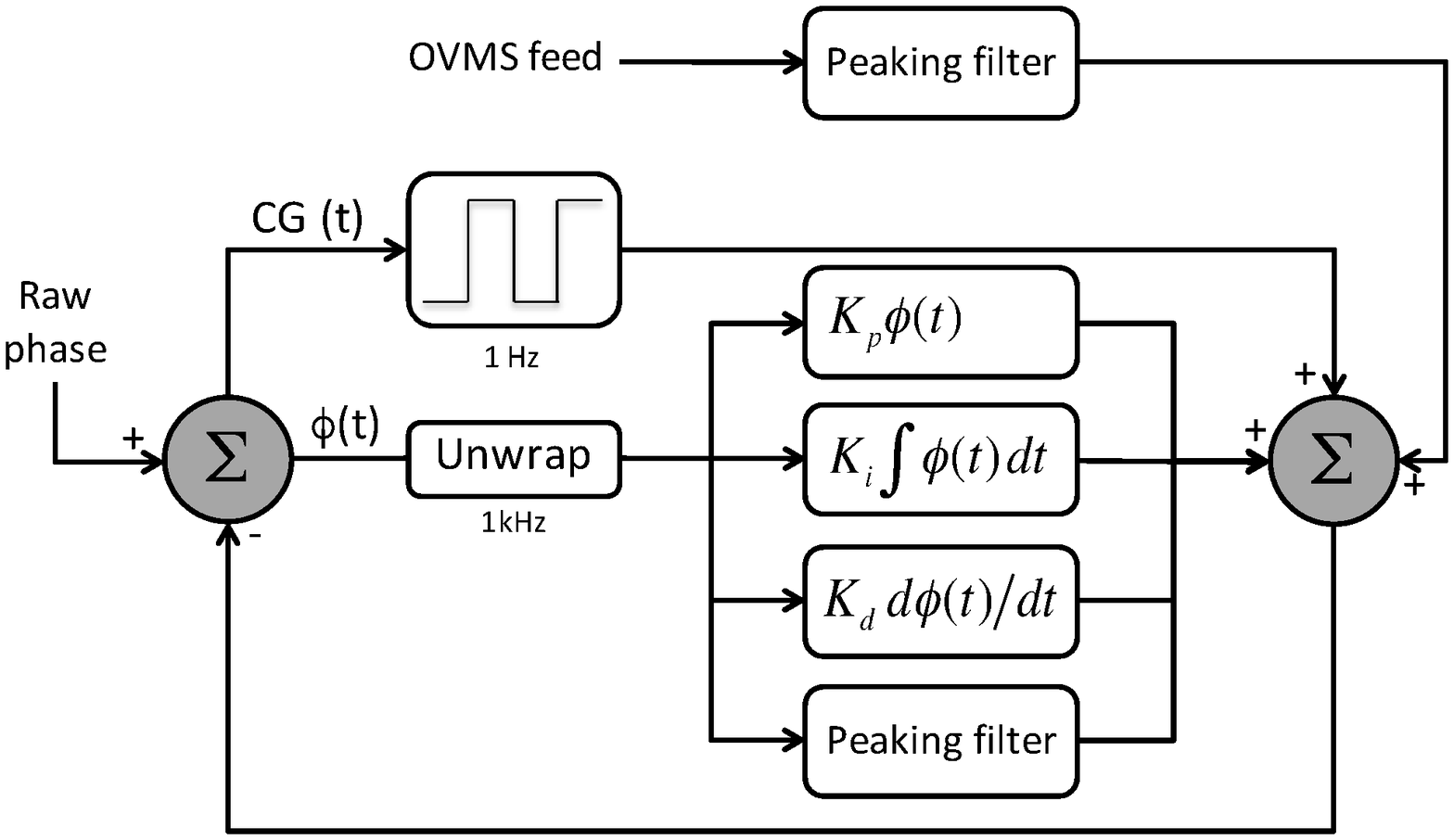}
		\vspace{0.2cm}
		\caption{{\bf Left}: Conceptual schematic of the nulling and PHASECam beam combination. Beam combination is done in the pupil plane on a 50/50 beamsplitter (BS), which can be translated to equalize the pathlengths between the two sides of the interferometer. To achieve an achromatic suppression of light over a sufficiently large bandwidth (8-13\,$\upmu$m), a compensator window (CW) with a suitable thickness of dielectric is introduced in one beam. Both outputs of the interferometer are directed to the near-infrared phase sensor (PHASECam) while one output is reflected to the NOMIC science detector with a short-pass dichroic. Note that this sketch does not show several fold mirrors and biconics\cite{Hinz:2008}. {\bf Right}: Block diagram of LBTI OPD controller. The measured phase is first unwrapped and then goes through a classical PID controller. A peaking filter is also used to improve the rejection of specific vibration frequencies. An outer loop running at typically 1\,Hz is used to monitor the group delay and capture occasional fringe jumps. In addition, real-time OPD variations induced by the LBT structure are measured by accelerometers all over the telescope (OVMS system) and fedforward to LBTI's path length corrector.}\label{fig:combination}
\end{center}
\end{figure}

\begin{figure}[!t]
	\begin{center}
		\includegraphics[height=5.5 cm]{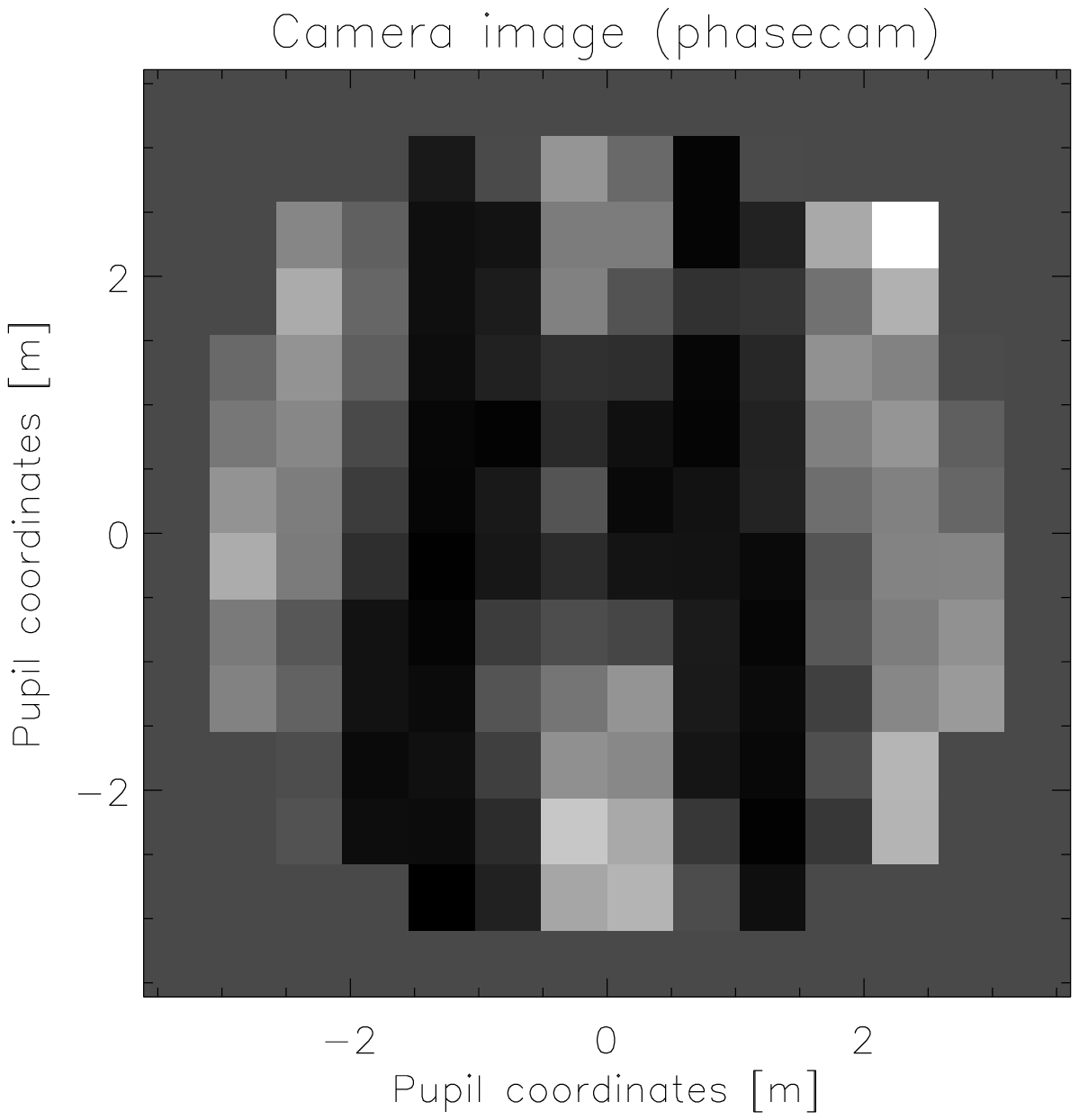}
		\includegraphics[height=5.5 cm]{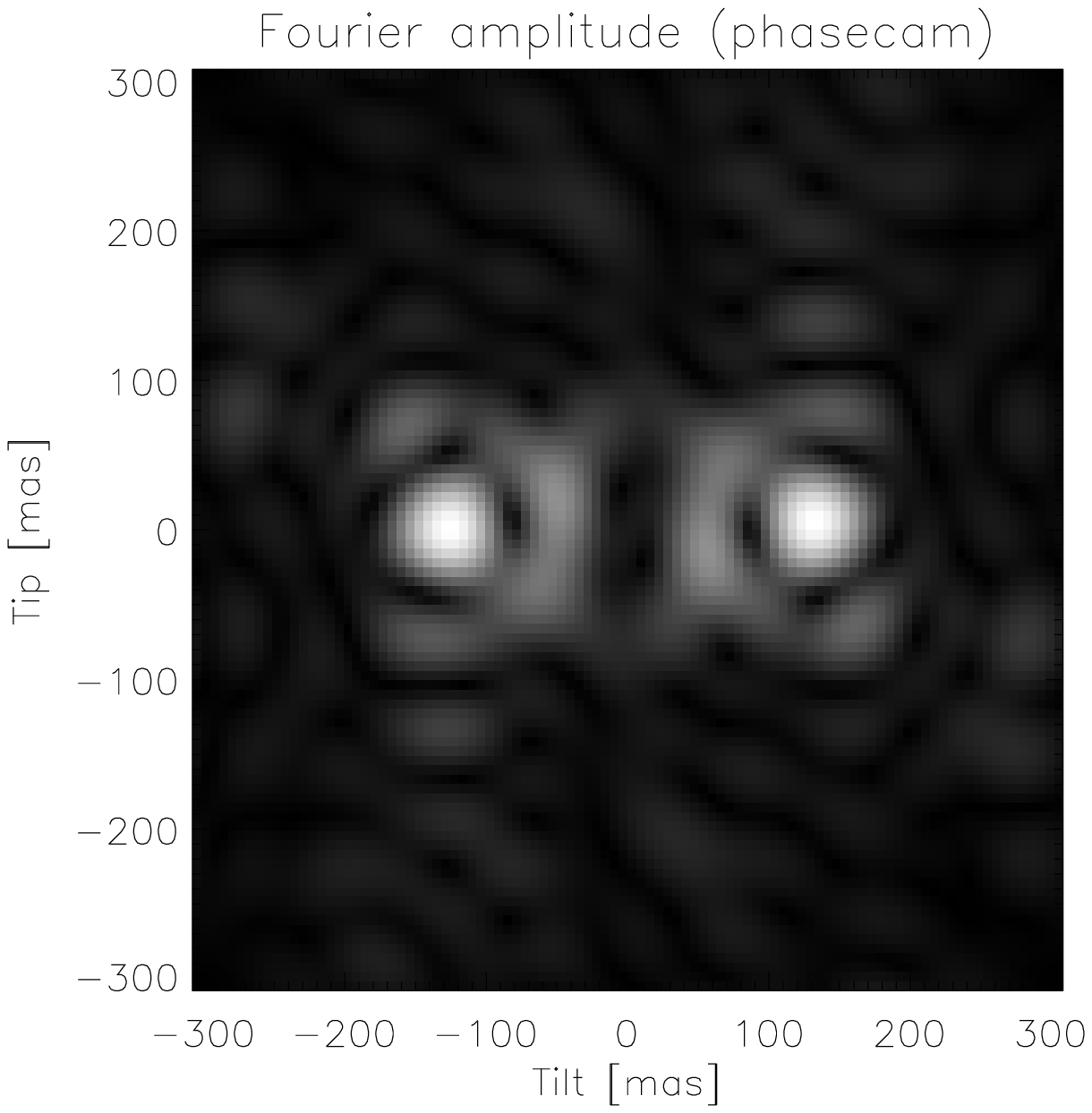}
		\includegraphics[height=5.5 cm]{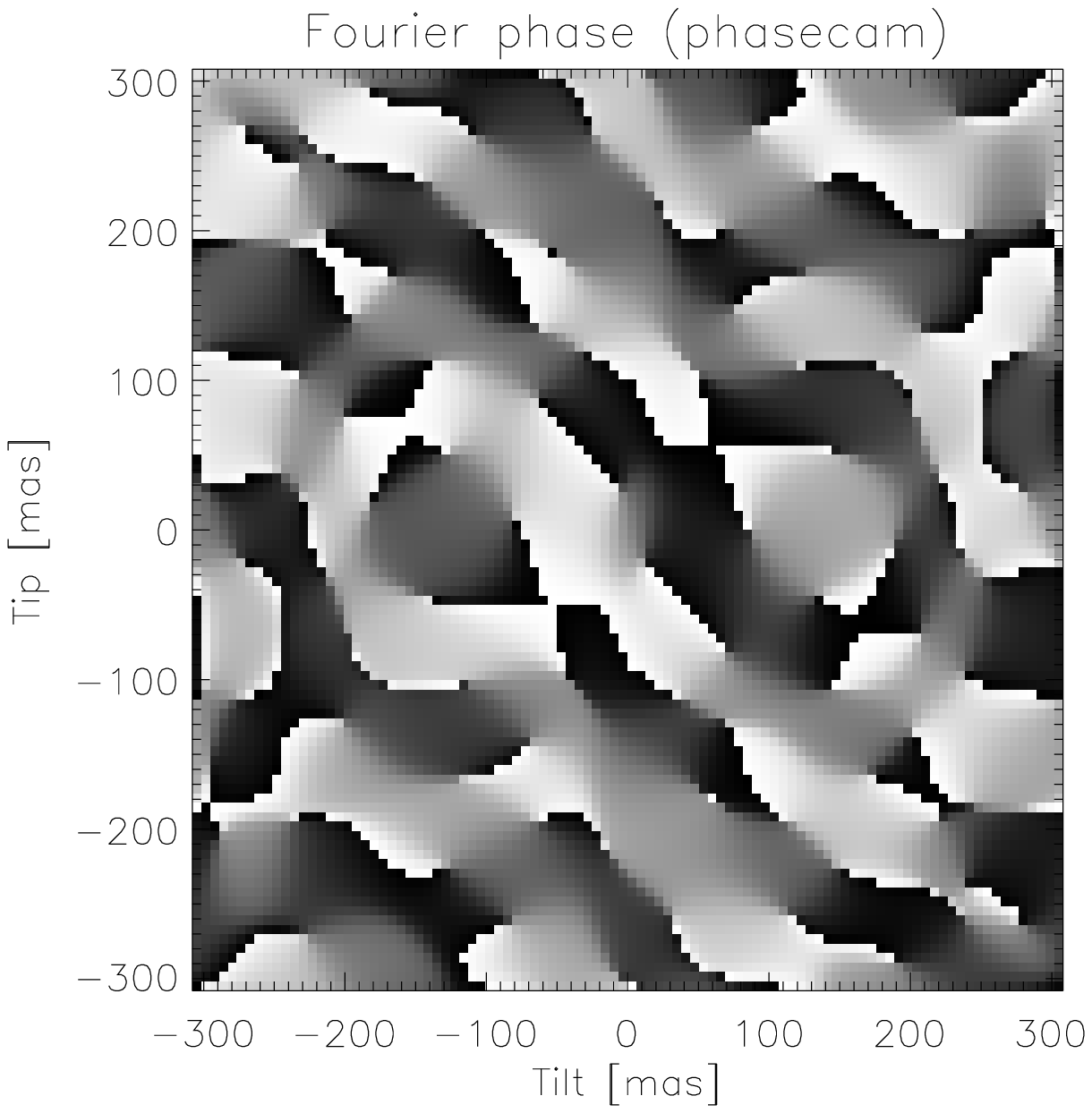}
		\caption{LBTI's phase sensing approach. Pupil images of two interferometric outputs are formed on PHASECam (one output shown on the left) and the Fourier transform is computed to sense both differential tip/tilt and OPD. The peak position in the amplitude of the Fourier transform (middle image) gives a measurement of the differential tip/tilt while the argument of the Fourier transform (right image) at the peak position gives a measurement of the optical path delay.} \label{fig:approach}
	\end{center}
\end{figure}

Differential tip-tilt and phase variations between the two AO-corrected apertures are measured with PHASECam\cite{Hill:2006,Hinz:2008}, LBTI's near-infrared camera. PHASECam uses a fast-readout PICNIC detector that receives the near-infrared light from both interferometric outputs (see left panel of Figure~\ref{fig:combination}). Various sensing approaches are implemented and the default mode consists in forming a K-band image of pupil fringes (equivalent to wedge fringes) on the detector and measure both the phase and the differential tip/tilt simultaneously via Fourier transform (see Figure~\ref{fig:approach}). Because of the nature of the Fourier transform measurement, the phase measurement is limited to phase values in the  [$-\pi$,$\pi$] range so that any phase error larger than $\pi$ is not detected and, hence, not corrected. On-sky verification showed that such phase jumps occur occasionally even with the phase loop closed at 1\,kHz. In order to get around this issue, the envelope of interference (or the group delay) is tracked simultaneously via the change in contrast of the fringes (i.e., contrast gradient or CG\cite{Defrere:2014}). This measurement is done at slower speed, typically 1\,Hz, and is used to detect and correct any phase jumps not captured by the Fourier transform phase measurement. Actually, this slow drift between group delay and phase delay is exactly the effect expected from water vapor turbulence.

In addition to the real-time control described above, OPD and individual tip/tilt motions induced by the telescope structure are fedforward to a fast pathlength corrector (FPC) using real-time data from 30 active accelerometers all over the telescope (Optical Vibration Monitoring System, OVMS\cite{Kurster:2010}). A new software system, called OVMS$^+$, was recently tested and provided significant improvements in the loop stability and performance\cite{Bohm:2016}. Peaking filters are also used to improve the rejection of specific vibration frequencies not captured by the feedfoward system. They are biquad filters applied directly to the control algorithm that drives the FPC. Commissioning observations have shown that the loop can run at full speed down to a magnitude of K$\sim$6.5, which is sufficient to observe all targets of the LBTI exozodiacal dust survey (Hunt for Observable Signatures of Terrestrial Planetary Systems, HOSTS\cite{Danchi:2014}). The loop frequency can in principle be decreased down to $\sim$30Hz to observe fainter objects but this has never been tested and requires more investigation. 

\subsection{Water vapor measurements}

The sensing approach described in the previous section is performed at K band whereas the science channels of the LBTI generally operate at longer wavelengths. In order to track the phase variations induced by water vapor seeing, the usual method at other facilities is to use the wavelength dependence of the phase in the bandpass of the fringe sensor (e.g., method used for the KIN\cite{Colavita:2010}). In the case of the LBTI, we have added new filters to the PHASECam optics in order to use a two color sensing approach. The measurement approach is then similar to that used for the KIN, the only difference being that we use the phase measured in two broadband channels (H and K) rather than the curvature of the phase within the K band. One output of the interferometer (see left part of Figure~\ref{fig:combination}) is now obtained at H band, while the main output remains at K band. The idea is to measure deviations from achromatic OPD variations using simultaneous H- and K-band unwrapped phase measurements. To be specific, the first step of the algorithm is to measure the difference in phase at H and K band (or the group delay):

\begin{equation}
\phi_{\rm gd} = \phi_{\rm H} -  \phi_{\rm K}\, ,
\end{equation}
where $\phi_{\rm K}$ and $\phi_{\rm H}$ are obtained separately from each output of the interferometer. Then, we construct a ``pseudo-K" phase term from the group delay as:
\begin{equation}
\phi_{\rm pseudo-K} = \phi_{\rm gd}\frac{\lambda_{\rm H}}{\lambda_{\rm H} -  \lambda_{\rm K}}\, ,
\end{equation}
which can understood as the K-band phase calculated from the H-band phase under achromatic assumption. Finally, we compute the so-called ``water term'' as the difference between the two phase values:
\begin{equation}
wt = \phi_{\rm pseudo-K} - \phi_{\rm K}\, .\label{eq:wt}
\end{equation}
This water term is used to estimate the differential water vapor columns above each aperture and predict phase variations at other wavelengths (see Section~\ref{sec:pwv_opd}). 

\begin{figure*}[!t]
	\begin{center}
		\includegraphics[height=6.4 cm]{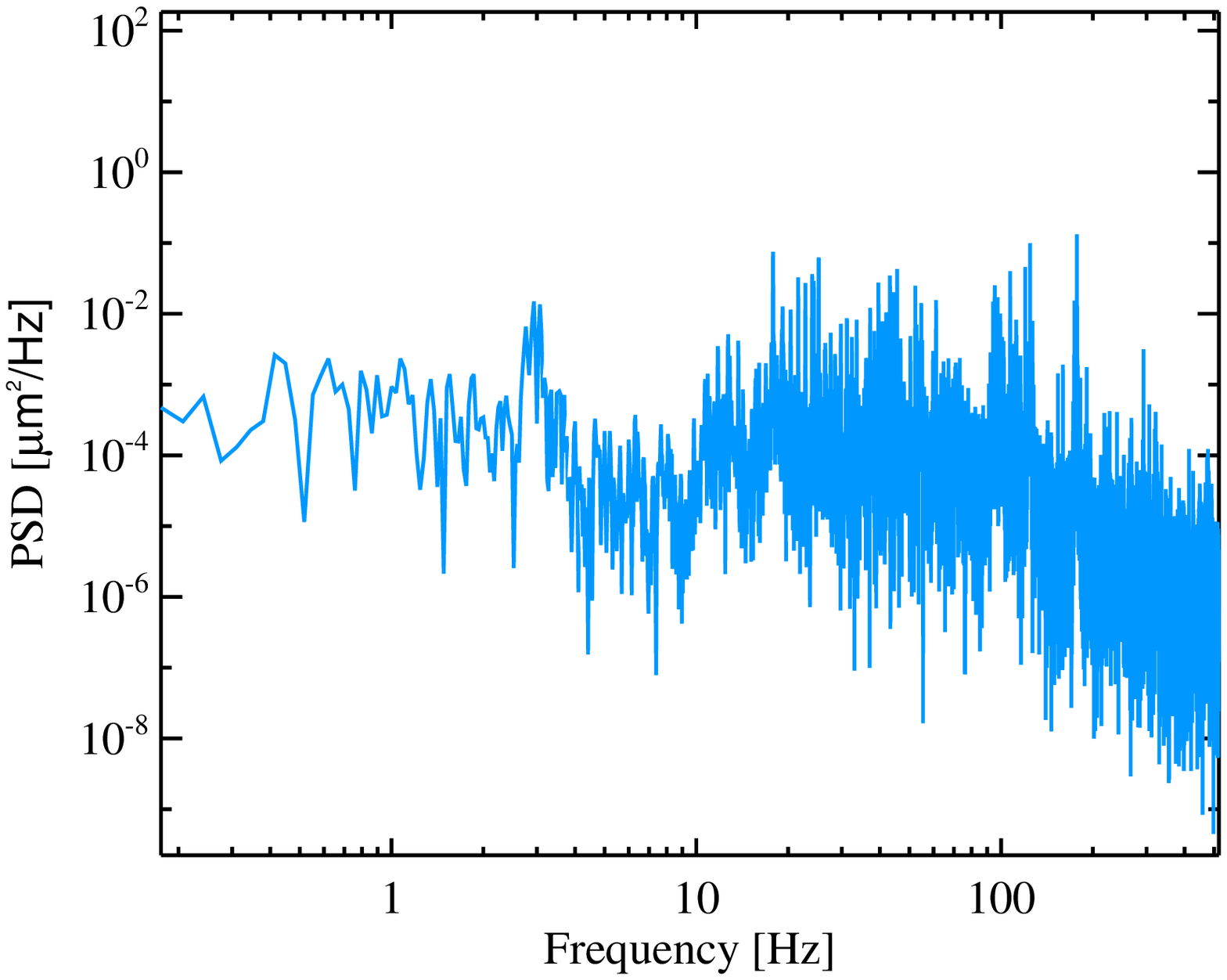}
		\includegraphics[height=6.4 cm]{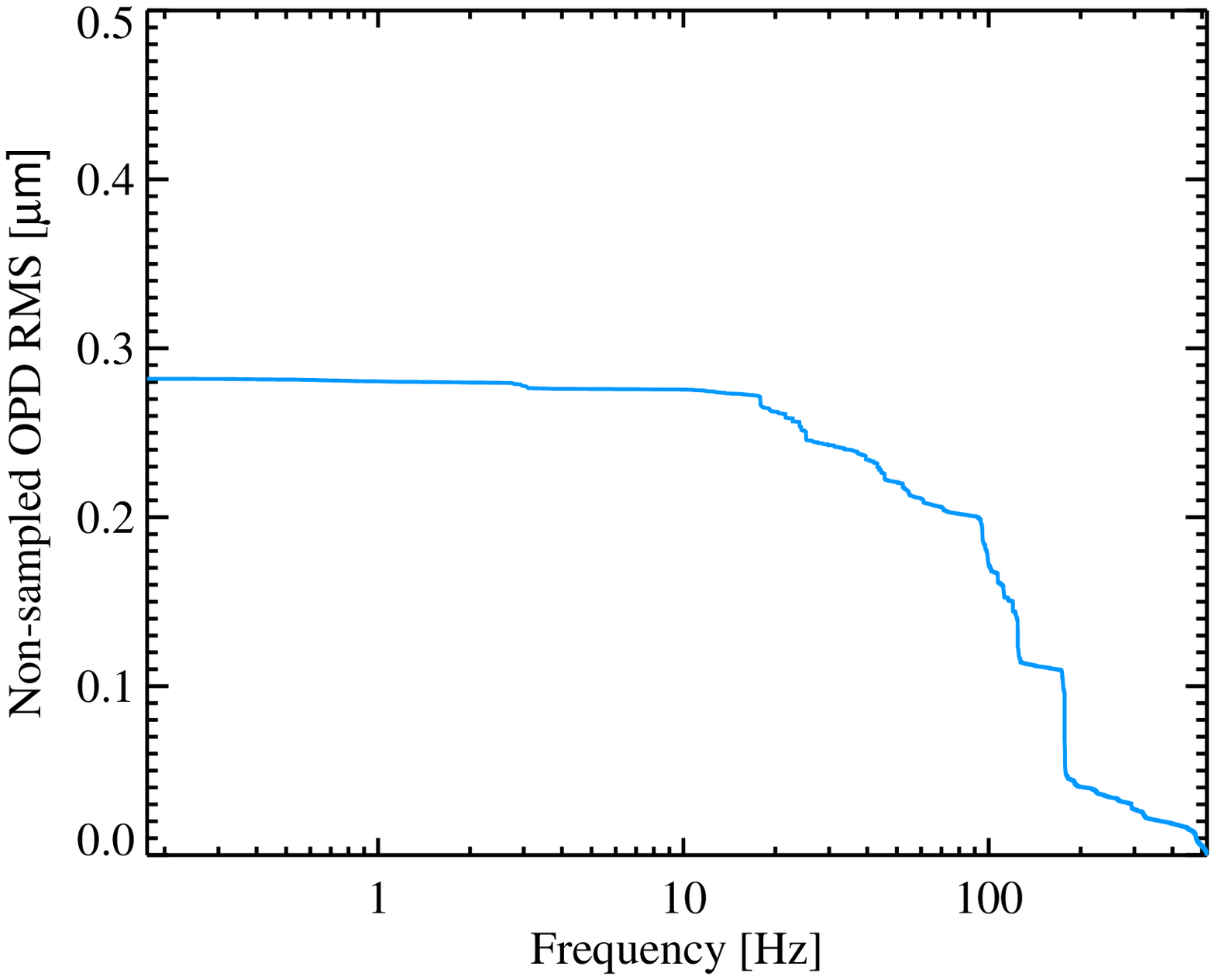}
		\caption{Left, typical closed-loop power spectral densities of the OPD variations between the two AO-corrected LBT apertures. Right, corresponding reverse cumulative OPD variations showing than most OPD residuals come from from high-frequency perturbations ($>20$\,Hz). Data were obtained on April 18, 2016 on the bright star HD163770 (K=1.0,V=3.9). Loop gains were $K_p$=0, $K_i$=300, and $K_d$=0 and feedforward using OVMS$^+$ activated (for OPD only and with no latency).}\label{fig:opd_perfo}
	\end{center}
\end{figure*}
\begin{figure*}[!t]
	\begin{center}
		\includegraphics[height=5.6 cm]{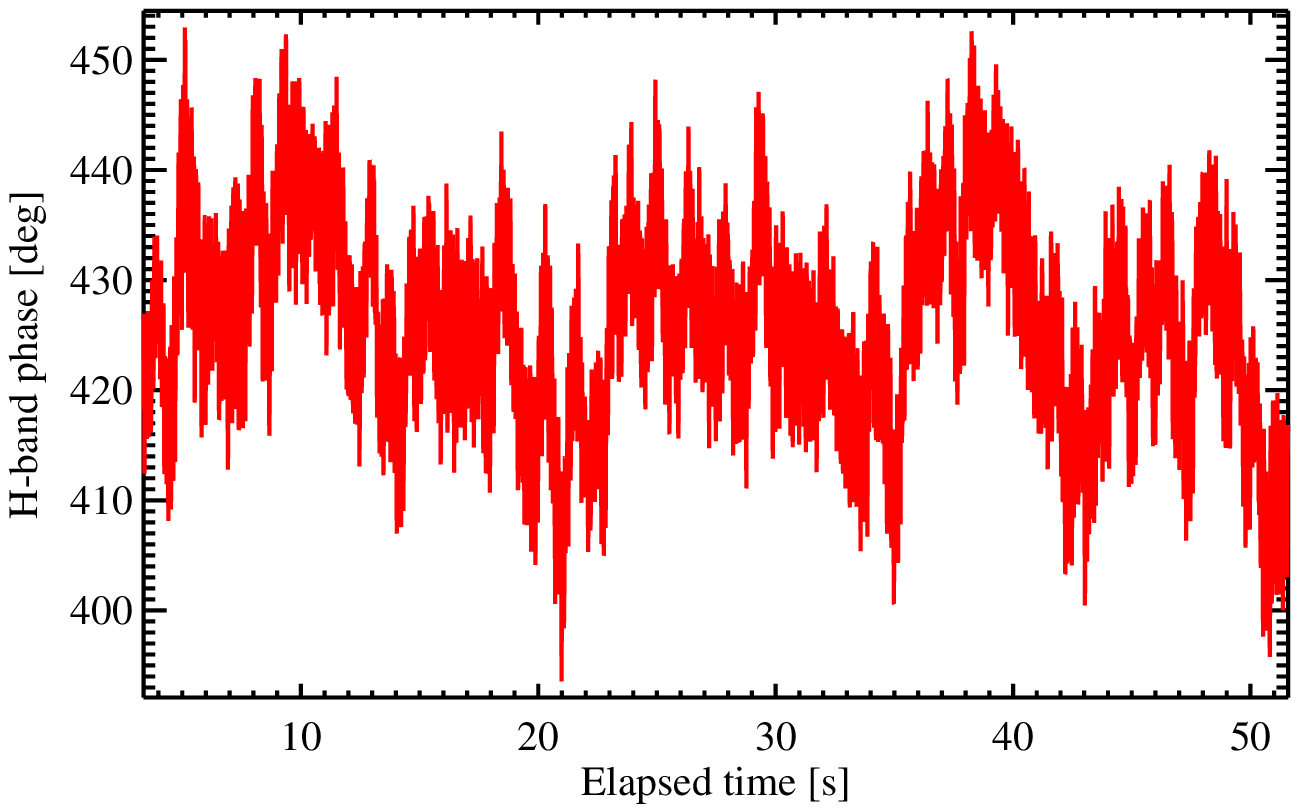}
		\includegraphics[height=5.6 cm]{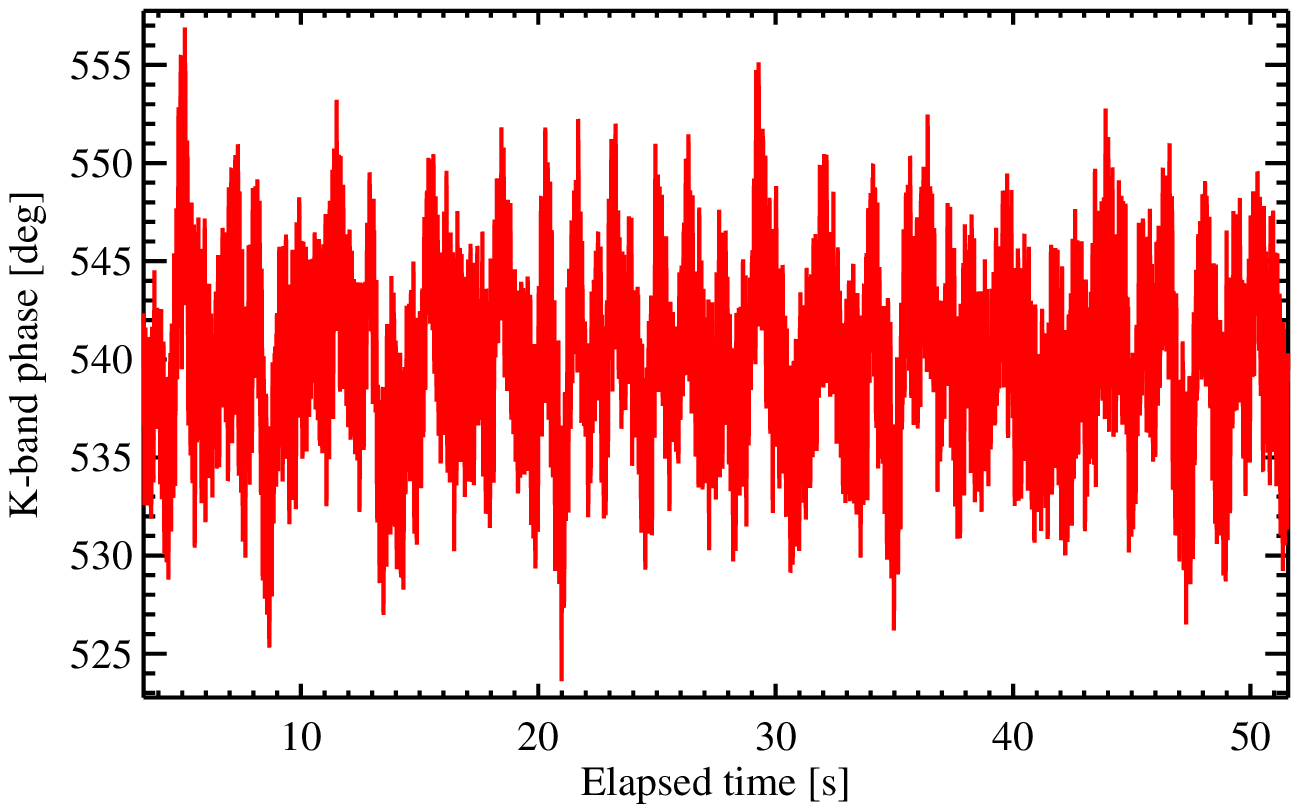}
		\includegraphics[height=5.6 cm]{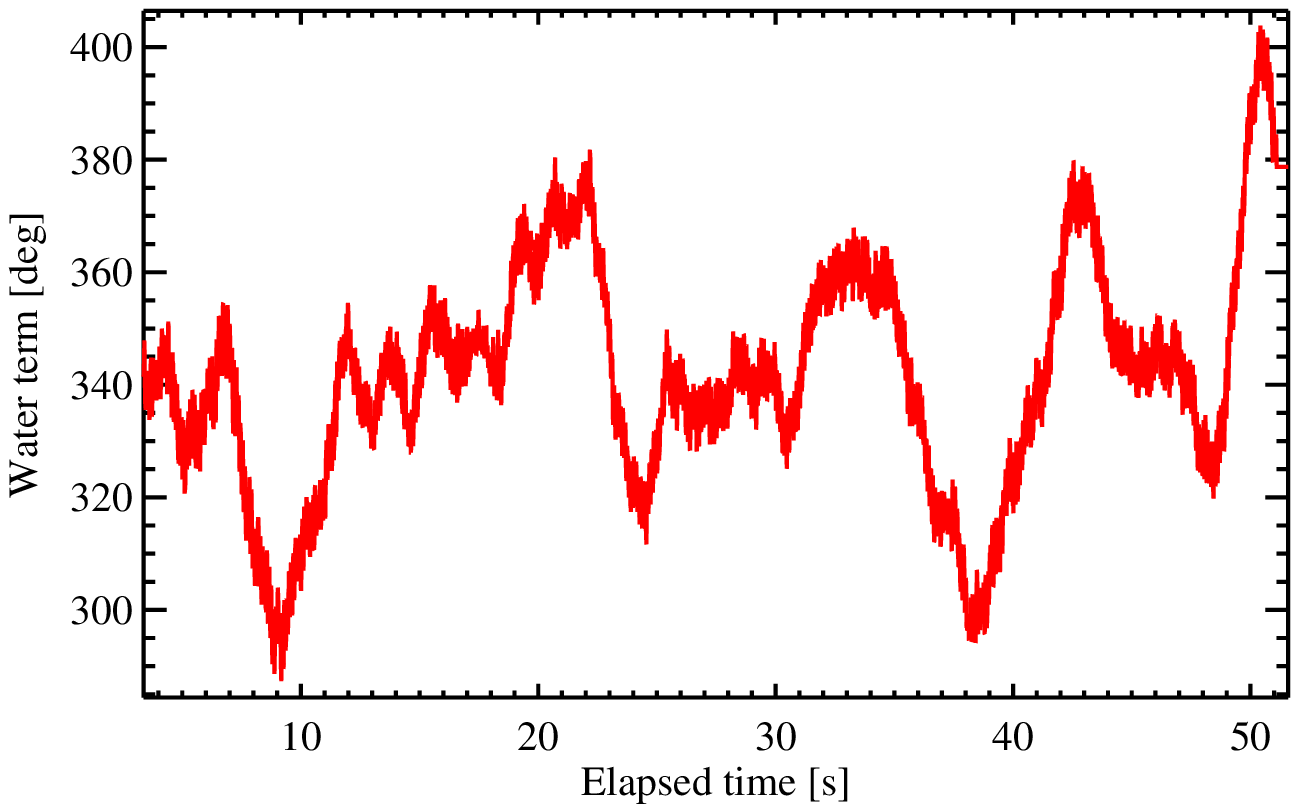}
		\includegraphics[height=5.6 cm]{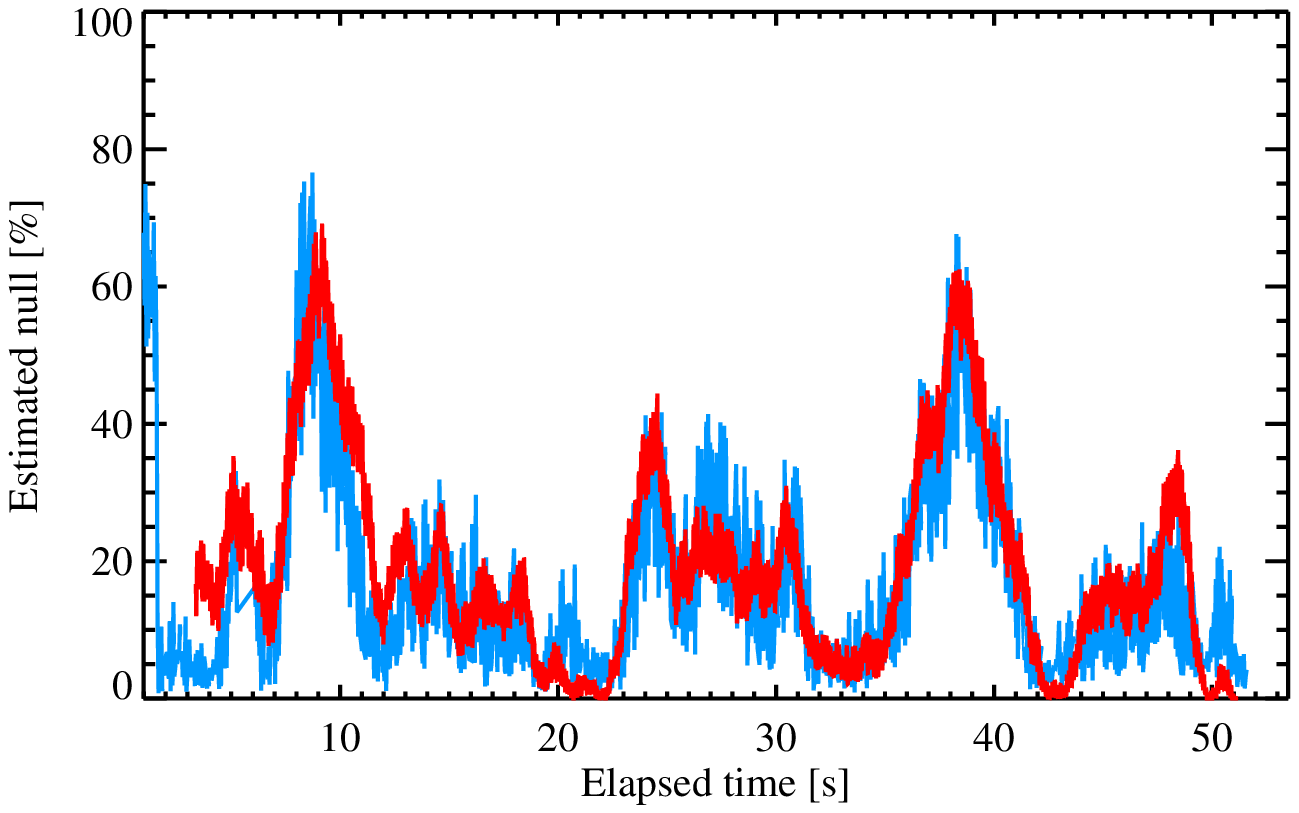}
		 \caption{Example of simultaneous H-band (top left) and K-band (top right) phase measurements obtained under wet conditions with PHASECam at 1\,kHz and boxcar-averaged over a period of 1 second to better show the low frequency fluctuations due to water vapor. The corresponding water term is shown in the bottom left plot while the simultaneous N'-band measurements are represented in the bottom right plot (blue line). The null estimated from the water term is over-plotted in red and shows a good correspondance with the actual null measurements. }\label{fig:wt}
	\end{center}
\end{figure*}

\section{RESULTS}
\label{sec:measurement}  

\subsection{Dry air OPD}\label{sec:phase}

Figure~\ref{fig:opd_perfo} (left) shows the closed-loop power spectral density (PSD) of the OPD variations obtained in April 2016 under typical observing conditions. As expected, the low-frequency ($<$20-30\,Hz) component of the seeing is strongly removed and the OPD residuals are dominated by high-frequency perturbations due to resonant optics inside the LBTI cryostat. The closed-loop residual OPD currently amounts to approximately 280\,nm RMS as shown by the right part of Figure~\ref{fig:opd_perfo}. This is approximately 150\,nm better than previously-published results obtained one year ago\cite{Defrere:2016}. This improvement is mostly due to an improved real-time control loop, a better vibration environment, and the routine use of the OVMS+ software\cite{Bohm:2016} for feedforward.

\subsection{Water vapor OPD}\label{sec:pwv_opd}

Figure~\ref{fig:wt} shows an example of simultaneous H-band (top left) and K-band (top right) phase measurements obtained under wet conditions. The data have been obtained at 1\,kHz and boxcar averaged in order to better show the low frequency fluctuations due to water vapor. As expected, the phase RMS measured at H band (9 degrees or 41nm) is larger than that measured at K band (4.5 degrees or 28nm) due to variable atmospheric dispersion. Combining the two terms using Equation~\ref{eq:wt}, we can form the water term represented in the bottom left panel of Figure~\ref{fig:wt} and predict the phase fluctuations in the bandpass of the nuller. Using this predicted phase, we can construct the expected null variations from the water term. This is shown in the bottom right part of Figure~\ref{fig:wt}  by the red line  computed using a multiplicative gain on the water term of 6 and a constant phase offset of -240 degrees. Simultaneous null measurements are over-plotted in blue and demonstrate the feasibility of our approach. The next step will consist in using this signal to adjust the null setpoint in real time and, therefore, mitigate the phase variations in the nuller bandpass.

%\section{IMPACT ON NULLING INTERFEROMETRY}
%\label{sec:nulling}  

\section{SUMMARY AND FUTURE WORK}
\label{sec:summary} 

Recent efforts in co-phasing the LBTI were focused on two main areas: (1) improving the performance of the system at K band where the fringe sensor operates and (2) compensating for atmospheric dispersion with the implementation of a new sensing mode. Thanks to an improved real-time control loop, a better vibration environment, and the routine use of OPD feedforward, the LBTI currently achieves a co-phasing stability of 280\,nm RMS at 1\.kHz. This performance is currently limited by high-frequency phase fluctuations occurring inside the LBTI cryostat and future work will focus on identifying and damping the source of these vibrations. Regarding atmospheric dispersion, a new mode was implemented to measure the phase at H and K band simultaneously (each using a different output of the interferometer). Using the information provided by the two wavelengths, it is possible to predict the phase variations at N band where the LBTI nuller operates. We have recently demonstrated the feasibility of this approach with simultaneous H-, K-, and N-band data. Future work will focus on using this signal to adjust the null setpoint in real time and, therefore, mitigate the phase variations in the nuller bandpass.

%%%%%%%%%%%%%%%%%%%%%%%%%%%%%%%%%%%%%%%%%%%%%%%%%%%%%%%%%%%%%
\acknowledgments     %>>>> equivalent to \section*{ACKNOWLEDGMENTS}       
The authors are grateful to M.~Colavita, O.~Absil, A.~Matter, and J.~Meisner for helpful discussions on fringe sensing and water vapor OPD measurements. LBTI is funded by a NASA grant in support of the Exoplanet Exploration Program. The LBT is an international collaboration among institutions in the United States, Italy and Germany. LBT Corporation partners are: The University of Arizona on behalf of the Arizona university system; Istituto Nazionale di Astrofisica, Italy; LBT Beteiligungsgesellschaft, Germany, representing the Max-Planck Society, the Astrophysical Institute Potsdam, and Heidelberg University; The Ohio State University, and The Research Corporation, on behalf of The University of Notre Dame, University of Minnesota and University of Virginia. 
%%%%%%%%%%%%%%%%%%%%%%%%%%%%%%%%%%%%%%%%%%%%%%%%%%%%%%%%%%%%%

\bibliographystyle{spiebib}   %>>>> makes bibtex use spiebib.bst

\end{document}